\documentclass[a4paper,11pt]{amsart}
\begin{document}

\title[On the gravitational collapse of a massive star]
{{\bf On the gravitational collapse
\\ of a massive star}}
\author{Angelo Loinger}
\date{}
\address{Dipartimento di Fisica, Universit\`a di Milano, Via
Celoria, 16 - 20133 Milano (Italy)}
\email{angelo.loinger@mi.infn.it}
\author{Tiziana Marsico}
\date{}
\address{Liceo Classico ``G. Berchet'', Via della Commenda, 26 - 20122
                                           Milano (Italy)}
\email{martiz64@libero.it}
\thanks{To be published on \emph{Spacetime \& Substance}}

\begin{abstract}
The celebrated treatment of continued gravitational collapse by
Oppenheimer and Snyder is revisited and emended from some inherent
flaws. The star contracts itself into a material point, not into a
black hole.
\end{abstract}

\maketitle

\vskip1.20cm
% section{}
\noindent {\bf 1.}-- The gravitational spherically-symmetric
collapse of a massive (i.e., of mass greater than a few solar
masses) star, simply described as ``dust" cluster of a spherical
shape (with negligible pressure), has been firstly investigated in
1939 by Oppenheimer and Snyder \cite{1}, whose approach is now a
standard reference \cite{2}. However, treatments \cite{1} and
\cite{2} are not exempt from conceptual flaws, and therefore the
problem deserves a re-examination.

\vskip0.80cm
%\section{}
\noindent {\bf 2.}-- The above mentioned collapse is described
with the same mathematical formalism of one of Friedmann's
cosmological models, and precisely the oscillating model in its
contraction phase. Friedmann's universe -- as it is well known --
consists of a spherical ``dust" cluster of ``point" galaxies
interacting only gravitationally, and it is based on the principle
of homogeneity and isotropy of the three-dimensional space,
according to which at any instant of time it is seen similarly by
all galactic observers \cite{3}.
\par
First of all, we remember that for a dust cluster of particles,
which interact \emph{only} gravitationally, \emph{the world lines
of the particles are geodesic lines}.
\par
Quite generally, in a Gaussian-normal (or synchronous) frame of
reference -- for which

\begin{equation} \label{eq:one}
g_{00}=1\quad {;}\quad g_{0\alpha}=0 \quad{;} \quad (\alpha=1,2,3)
\quad {\rm -},
\end{equation}

the time lines are geodesic lines. Quite generally, in a co-moving
frame of reference the world lines coincide with the  the time
lines.
\par
Accordingly, if we choose for our dust cluster (our star) a
Gaussian-normal frame $S$, the time lines of the particles are
both geodesic and world lines -- and the frame $S$ is also a
co-moving one. We have:

\begin{equation} \label{eq:two}
\frac{\textrm{d}x^0}{\textrm{d}s} = 1\quad {\rm ;} \quad
\frac{\textrm{d}x^\alpha}{\textrm{d}s}=0 \quad.
\end{equation}

The matter tensor of the dust is:

\begin{equation} \label{eq:three}
T^{jk}=\rho\frac{\textrm{d}x^j}{\textrm{d}s}\frac{\textrm{d}x^k}{\textrm{d}s}
\quad, \quad (j,k=0,1,2,3) \quad, \qquad (c=1) \quad,
\end{equation}

where $\rho$ is the invariant mass density. In the above frame $S$
it is simply:

\begin{equation} \label{eq:four}
T_{00}=T^{00}=\rho \quad ; \qquad \quad
T^{0\alpha}=T^{\alpha\beta}=0 \quad.
\end{equation}

The $\textrm{d}s^{2}$ of our problem is given by
Friedmann-Robertson-Walker metric:

\begin{equation} \label{eq:five}
\textrm{d}s^2=(\textrm{d}x^0)^2-A^2(r^2)F^2(x^0)
\left[(\textrm{d}x^1)^2+(\textrm{d}x^2)^2+(\textrm{d}x^3)^2\right]
\quad,
\end{equation}

where:

\begin{equation} \label{eq:six}
A(r^2) \equiv \left[1+\frac{r^2}{4}\right]^{-1} \quad; \qquad
r^2=(x^1)^2+(x^2)^2+(x^3)^2 \quad,
\end{equation}

and $F(x^0)$ must be determined by Einstein field equations:

\begin{equation} \label{eq:seven}
R_{jk}-\frac{1}{2}g_{jk}R=-\kappa T_{jk} \quad, \qquad (\kappa
\equiv 8\pi G) \quad,
\end{equation}

from which:

\begin{equation} \label{eq:eight}
\frac{1}{F} + \frac{\dot{F}^2}{F} +2\ddot{F}=0 \quad,
\end{equation}

\begin{equation} \label{eq:nine}
\frac{1}{F^2} + \frac{\dot{F}^2}{F^2} - \frac{1}{3}\kappa\rho=0
\quad.
\end{equation}

Their solution describes a periodic oscillation between $F=0$ and
a given maximum value $F_{\textrm{max}}$. For $F=0$ the density
$\rho$ is infinite and the gravitational field is singular.
Starting from $F_{\textrm{max}}$ our star contracts itself, in a
finite time, into a \emph{material point} corresponding to $F=0$.

\vskip1.20cm
% section{}
\noindent {\bf 3.}-- In the Newtonian model analogous to the
relativistic model of sect.\textbf{2}., a particle on the
spherical surface of radius $\Re(t)$ of the star is attracted by
the mass $M$ within the sphere according to Newton's law:

\begin{equation} \label{eq:ten}
\ddot{\Re} = - \frac{\kappa M}{8\pi \Re^2} \quad, \qquad
\textrm{with} \quad M=\frac{4}{3} \pi \Re^{3}(t)
\rho(t)=\textrm{constant \quad ;}
\end{equation}

thus

\begin{equation} \label{eq:eleven}
\ddot{\Re} + \frac{\kappa}{6} \Re \rho = 0 \quad,
\end{equation}

from which

\begin{equation} \label{eq:twelve}
\frac{1}{\Re^2}+\frac{\dot{\Re}^2}{\Re^2} -
\frac{1}{3}\kappa\rho=0 \quad.
\end{equation}

Therefore we can write the following equations, which are formally
\emph{identical} to eqs. (\ref{eq:eight}) and
(\ref{eq:nine}):

\begin{equation} \label{eq:thirteen}
\frac{1}{\Re} + \frac{\dot{\Re}^2}{\Re} +2\ddot{\Re}=0
\end{equation}

\begin{equation} \label{eq:fourteen}
\frac{1}{\Re^2}+\frac{\dot{\Re}^2}{\Re^2} -
\frac{1}{3}\kappa\rho=0 \quad.
\end{equation}

They give a periodic oscillation between $\Re=0$ and a given
$\Re_{\textrm{max}}$. Starting from $\Re_{\textrm{max}}$ the star
contracts itself into a \emph{material point} (for which
$\rho=\infty$) corresponding to $\Re=0$.
\par
Remark, in conclusion, that: \emph{i}) in both cases --
Einsteinian Friedmann's model and Newtonian model -- we have
employed only \emph{one} co-ordinate frame; \emph{ii}) the
gravitational field \emph{outside} the star has caused \emph{no}
problem, and it has been ignored.

\vskip1.20cm
% section{}
\noindent {\bf 4.}-- The approach by Oppenheimer and Snyder
(\cite{1}, \cite{2}) can be criticized for the following reasons:
\emph{i}) the consideration of the Einsteinian field
\emph{outside} the star is a useless superfetation, because the
Gaussian-normal frame of FRW-metric characterizes exhaustively the
phenomenon of collapse; \emph{ii}) in \cite{1}, \cite{2} the
external gravitational field is described by the
\emph{\textbf{standard}} form of solution of the Schwarzschild
problem -- \emph{erroneously} called ``Schwarzschild solution"
\cite{4} --, which is \emph{properly} valid only for the values of
radial co-ordinate $r$ greater than $\kappa M/(4\pi)$.
\par
At any rate, if one wishes to take into consideration also the
outside field, it is suitable to proceed as follows.
\par
As it is well known, the necessary and sufficient condition that a
Riemann-Einstein manifold admit the group of spatial rotations is
that its $\textrm{d}s^2$ be reducible to the following form, which
holds both internally and externally to the matter distribution:

\begin{equation} \label{eq:fifteen}
\textrm{d}s^2=B_1(r,t)\textrm{d}t^2-B_2(r,t)\textrm{d}r^2-r^2
\textrm{d}\omega^2 \quad, \qquad( r>0; c=1) \quad,
\end{equation}

where

\begin{equation} \label{eq:sixteen}
\textrm{d}\omega^2 \equiv \textrm{d}\theta^2+\sin^2\theta
\textrm{d}\phi^2 \quad,\qquad(0\leq\theta<\pi;\quad
0\leq\phi<2\pi) \quad.
\end{equation}

By virtue of a famous Birkhoff 's theorem, the outside field is
\emph{time-independent}, and consequently the general form of
\emph{external} metric is given by the following formula, as it
was emphasized by Eddington \cite{5}:

\begin{equation} \label{eq:seventeen}
\textrm{d}s^2=\left(1-\frac{2m}{f(r)}\right)\textrm{d}t^2-\left(1-\frac{2m}{f(r)}\right)^{-1}
[\textrm{d}f(r)]^2-[f(r)]^2\textrm{d}\omega^2 \quad;
\end{equation}

here: $m\equiv\kappa M/(8\pi)$, and $f(r)$ is \emph{any} regular
function of $r$. \emph{\textbf{No physical result depends on the
choice of} $f(r)$}.
\par
If we put

\begin{equation} \label{eq:eighteen}
f(r)\equiv r + 2m \quad,
\end{equation}

we obtain the form of solution first investigated by M. Brillouin
\cite{6}, which holds for $r>0$. Putting

\begin{equation} \label{eq:nineteen}
f(r)\equiv \left[r^3+(2m)^3\right]^\frac{1}{3} \quad ,
\end{equation}

we obtain the \emph{\textbf{original}} form of solution given by
Schwarzschild \cite{7}, which is valid for $r>0$; thus,
Schwarzschild's and Brillouin's forms are \emph{maximally
extended}. If we put simply
\begin{equation} \label{eq:twenty}
f(r)\equiv r \quad,
\end{equation}

we obtain the \emph{\textbf{standard}}, or Hilbert-Droste-Weyl,
form of solution, which holds only for $r>2m$. Remark that
Brillouin's form and Schwarzschild's original form are
diffeomorphic to the exterior part $r>2m$ of HDW-form. Within the
singular locus $r=2m$ of this last form the time co-ordinate $t$
takes the role of the radial co-ordinate $r$, and
\emph{vice-versa}. The solution loses its physical
\emph{Eigentlichkeit} (``appropriateness", according to Hilbertian
terminology), and becomes further \emph{non}-static. Consequently,
the notion of black hole -- the ``globe" $r=2m$ -- is destitute of
any meaning. As \emph{all} the Fathers of Relativity perfectly
knew!

\noindent{\emph{\textbf{Conclusion}}}. Take ideally an
instantaneous photograph of the contracting sphere at \emph{any}
time $\overline{t}$, and call $\overline{r}$ its co-ordinate
radius at this time. If we choose the regular function $f(r)$ of
Eddington's formula (\ref{eq:seventeen}) so that the corresponding
form of solution be valid for $r>0$, we can say: since
$\overline{t}$ is just any time and $\overline{r}$ tends to zero,
at the final stage of its contraction our star will become a
\emph{\textbf{point}} mass. Thus, also with this \emph{schematic},
 model-independent, consideration we see that no celestial body
can convert itself into a black hole.

\vskip0.80cm
%\section{}
\noindent {\bf APPENDIX A} \par \vskip0.10cm From the
\emph{experimental} standpoint we can affirm that
\textbf{\textbf{\emph{no}}} black hole (BH) has ever been
detected. Indeed, an accurate scrutiny of the papers in which
observational discoveries of stellar-mass, or of supermassive,
BH's are cried up, shows that in reality the authors have
discovered only celestial bodies of large, or enormously large,
masses concentrated in very small volumes \cite{8}.
\par
It is interesting to read the comments of Wolfgang Kundt, a
renowned astrophysicist, concerning the BH's \cite{9}. He does not
criticize the current theoretical viewpoint on the question, but
limits himself  to significant remarks of the following kind.
\par
$\alpha$) Page 37 of \cite{9}: ``The critical mass [...] which
determines whether a star eventually turns into a white dwarf or
something more compact -- if such a mass is well defined -- is
[...] controversial. It should be consistent with (i) the
birthrate of white dwarfs, (ii) the birthrate of neutron stars,
(iii) the PN [\emph{planetary nebula}] rate, (iv) the SN
$[$\emph{supernova}$]$ rate, (v) the supernova remnant (SNR) rate,
and (vi) the \emph{initial mass function} (IMF) which counts the
number of stars as a function of their mass at birth. In view of
the many neutron stars in the Galaxy -- detected as pulsars,
binary X-ray sources, or even invisible (when screened, without
accreting) -- I favour a \emph{critical mass} of some 5 solar
masses (over larger values, like 8 solar masses). The bias would
become even more severe if a large number of massive stars would
end up as black holes (BHs); in my own judgment, none of the BH
candidates (BHCs) do involve BHs, rather they are neutron stars
surrounded by massive disks [...]: The proposed BHCs have too much
spectral and variability structure, reminiscent of a rotating
inclined magnet at their center [...]."
\par $\beta$) Page 81 of \cite{9}: ``Another disk peculiarity is expected at the
\emph{centers of galaxies} [...] The galaxy feeds an
\emph{active}, nuclear-burning nucleus, a \emph{burning disk}
$[$Kundt, 2000$]$. -- Instead, most of my colleagues prefer to
think of a supermassive black hole as the central engine of all
the \emph{active galactic nuclei} (AGN). They have not convinced
me, after more than 20 years. AGN activity requires a refilling
engine, with nuclear burning, magnetic reconnections, and
explosive ejections of the ashes [...]. Black-hole formation would
require distinctly higher mass concentrations than are even
reached in galactic nuclei, by a factor of $10^2$. The quasar
phenomenon is a simple consequence of a permanent inward galactic
mass flow, at an average of $<1$ (or $\sim 1$) solar mass per
year, which piles up at the center."

\par $\gamma$) Page 101 and 102 of \cite{9}: ``How about
stellar-mass holes? [...] Over 45 \emph{black-hole candidates}
have been proposed during the past 30 years from the class of
binary X-ray sources, both high-mass and low-mass -- among them
Cyg X-1 and A0620-00 -- on account of their large mass function,
absence of strict periodicities, and absence of type-I bursts
(understood as nuclear detonations at neutron-star surfaces). To
me, all of them look like neutron stars surrounded by massive (
$\approx$5 solar masses) accretion disks, because of their often
hard spectra (up into the $\gamma$-ray range), highly structured,
[...] and because of their indistinguishable further properties,
as a class, from all the established neutron-star binaries [...].
They just fill the gap between the high-mass and low-mass compact
binary systems. -- And the postulated \emph{supermassive black
holes} at the centers of (all the active) galaxies? They were once
believed to be required for energy reasons. The nuclear burning
(of H to Fe) is almost as efficient a lamp as black-hole
accretion, yielding a garanteed $<1\%$ (or $\sim 1\%$) of the rest
energy [...]. Besides,[...] the \emph{universality} of the
\emph{jet phenomenon} suggests a universal engine which we know is
a fast rotating magnet in the cases of newly forming stars, binary
neutron stars, and forming binary white dwarfs [Kundt, 1996,
2000]. -- [...] I share the doubts of a few other people, among
them (the late) Viktor Ambartsumyan and Hoyle et al. [2000], in
the widely accepted black-hole paradigm. [...]. Active galactic
nuclei may owe their extreme properties to those of their central
disks."

\par $\delta$) Pages 131 and 132 of \cite{9}: ``As already mentioned [...], [see
\emph{supra} in $\gamma$)], the list of stellar-mass
\emph{black-hole candidates} contains presently over 45 entries,
five of them with \emph{high-mass} companions ($>6$ (or $\sim 6$)
solar masses), the rest with \emph{low-mass} ones ($<2$ (or $\sim
2$) solar masses). Their defining property is a mass in excess of
3 solar masses of their compact component. All the high-mass BHCs
are persistent sources whereas most of the low-mass ones are
transient, with recurrence times of decades. Every year, two or
more X-ray novae joint the list. [...]. -- My suspicion of the BH
interpretation comes from (i) a number of spectral and lightcurve
properties which require a \emph{hard surface}, an \emph{oblique
magnetic dipole}, and two dense, \emph{interacting windzones};
(ii) the \emph{indistinguishability}, as a class, of the BHCs from
the neutron-star binaries in all properties other than their
inferred mass; and (iii) the missing \emph{intermediate-mass
systems} which should naturally evolve into neutron-star binaries
with massive disks. [...]. -- Among the long list of remarkable
properties in which the BHCs are indistinguishable, as a class,
from neutron-star binaries are (j) the presence of a \emph{third}
(\emph{precessional}) \emph{period} of several months, 294 d in
the case of Cyg X-1; (jj)  a hard-soft state \emph{spectral
bimodality}, pivoting around 6 keV, and extending up to MeV; (jjj)
their \emph{flickering}, expressed by their X-ray \emph{power
spectra} which range from mHz to $>$ kHz (or $\sim$ kHz) and show
various \emph{quasi-periods}, in particular of several $10^2$Hz,
up to $1.2$ kHz, reminiscent of innermost Kepler periods, of a
spin period, and/or of beat frequencies thereof; (jv) their
\emph{jet-formation} capability [...]." --

\vskip0.80cm
%\section{}
\noindent {\bf APPENDIX B} \par \vskip0.10cm Perhaps the belief of
most of the theoretical astrophysicists in the physical existence
of BH's begins now to get cracked. We report below the abstract of
a recent paper by Fr\o nsdal \cite{10}, an author who anticipated
in 1959 \cite{11} the celebrated result by Kruskal \cite{12} and
Szekeres \cite{13}, which contributed so much to the conviction of
the real existence of the BH's.
\par
Here is the mentioned abstract \cite{10}: ``This paper studies the
interpretation of physics near a Schwarzschild black hole. A
scenario for creation and growth is proposed that avoids the
conundrum of information loss. In this picture the horizon recedes
as it is approached and has no physical reality. Radiation is
likely to occur, but it cannot be predicted."
\par
(Of course, with the phrase ``Schwarzschild black hole" the author
denotes the fictive object derived from the current unphysical
interpretation of the part $r\leq2m$ of the standard HDW-form of
solution of Schwarzschild problem -- as we have previously
emphasized; and ``horizon" denotes in this interpretation the
singular locus $r=2m$.)
\par
Accordingly, the observational astrophysicists are warned: the
existence of the black holes cannot be detected -- exactly as the
existence of cosmic ether.


\begin{thebibliography}{99}
\bibitem{4}
The so-called ``Schwarzschild solution" is in reality due to
Hilbert, Droste, and Weyl.
\end{thebibliography}
\end{document}